\documentclass[twocolumn,epjc3]{svjour3}  

\usepackage{fix-cm}
\usepackage{mathptmx}      %
\usepackage{textcomp}
\usepackage{amssymb,amsmath,verbatim,mathtools,needspace,enumitem,etoolbox,graphicx,microtype,afterpage,bigints,gensymb,tabularx,xspace,siunitx}
\usepackage{aasmacros}
\usepackage{widetext}
\usepackage{graphicx}%
\usepackage{dcolumn}%
\usepackage{bm}%
\usepackage{lipsum}
\usepackage[dvipsnames, table]{xcolor}
\usepackage{physics}
\usepackage{soul}
\usepackage[numbers,sort&compress]{natbib}
\AtBeginEnvironment{thebibliography}{\small}
\usepackage{graphics}
\definecolor{linkcolor}{rgb}{0.0,0.3,0.5}
\usepackage[colorlinks = true,
            linkcolor = linkcolor,
            urlcolor  = linkcolor,
            citecolor = linkcolor,
            anchorcolor = linkcolor]{hyperref}
\usepackage{array}
\usepackage{orcidlink}
\usepackage[draft,defaultcolor=red, commandnameprefix=always,authormarkup=none]{changes}
\newcommand{\ssim}{\mathchar"5218\relax\,}

\definechangesauthor{RB}

\newcommand{\uvec}[1]{\bm{\hat{#1}}}

\newcommand{\bicocca}{Dipartimento di Fisica “G. Occhialini”, Università degli Studi di Milano-Bicocca, Piazza della Scienza 3, 20126 Milano, Italy}
\newcommand{\infn}{INFN, Sezione di Milano-Bicocca, Piazza della Scienza 3, 20126 Milano, Italy}
\newcommand{\bham}{Institute for Gravitational Wave Astronomy \& School of Physics and Astronomy, University of Birmingham, Birmingham, B15 2TT, UK}
\newcommand{\mpa}{Max-Planck-Institut f{\"u}r Astrophysik, Karl-Schwarzschild-Straße 1, 85741 Garching, Germany}
\newcommand{\lyon}{Universit\'e Lyon, Universit\'e Claude Bernard Lyon 1, CNRS,
IP2I Lyon / IN2P3, UMR 5822, F-69622 Villeurbanne, France}
\newcommand{\cambridge}{Institute of Astronomy, University of Cambridge, Madingley Road, Cambridge, CB3 0HA, UK}
\newcommand{\kavlicam}{Kavli Institute for Cosmology, University of Cambridge, Madingley Road, Cambridge, CB3 0HA, UK}
\newcommand{\damtp}{Department of Applied Mathematics and Theoretical Physics, Centre for Mathematical Sciences, University of Cambridge, Wilberforce Road, CB3 0WA, UK}

\AtBeginDocument{\RenewCommandCopy\qty\SI}
\ExplSyntaxOn
\msg_redirect_name:nnn { siunitx } { physics-pkg } { none }
\ExplSyntaxOff
\DeclareSIUnit\solarmass{\ensuremath{M_{\odot}}}
\DeclareSIUnit\parsec{pc}
\DeclareSIUnit\lightspeed{$c$}
\DeclareSIUnit\year{yr}
\DeclareSIUnit\arcsecond{as}
\DeclareSIUnit\astronomicalunit{AU}
\DeclareSIUnit\clight{\ensuremath c}
\journalname{Eur. Phys. J. C}
\begin{document}

\title{Test for LISA foreground Gaussianity and stationarity: galactic white-dwarf binaries}

\institute{
\bicocca\label{bic}
\and 
\infn\label{infn}
\and 
\mpa\label{mpa}
\and 
\bham\label{bham}
\and 
\lyon\label{lyon}
\and 
\cambridge\label{cam}
\and 
\kavlicam\label{kav}
\and 
\damtp\label{dam}
}
\thankstext{e1}{e-mail: riccardo.buscicchio@unimib.it}

\author{Riccardo~Buscicchio\thanksref{bic,infn,bham}~\orcidlink{0000-0002-7387-6754}%
\and 
Antoine~Klein\thanksref{bham}~\orcidlink{0000-0001-5438-9152}
\and
Valeriya~Korol\thanksref{mpa,bham}~\orcidlink{0000-0002-6725-5935}
\and 
Francesco~Di~Renzo\thanksref{lyon}~\orcidlink{0000-0002-5447-3810}
\and 
Christopher~J.~Moore\thanksref{cam,kav,dam}~\orcidlink{0000-0002-5447-3810}
\and 
Davide~Gerosa\thanksref{bic,infn}~\orcidlink{0000-0002-0933-3579}
\and 
Alessandro~Carzaniga\thanksref{bic}~\orcidlink{0009-0009-4336-7412}
} 

\date{Received: 24 May 2025 / Accepted: 6 August 2025 / Published 20 August 2025 }

\maketitle

\abstract{
Upcoming space-based gravitational-wave detectors will be sensitive to millions and resolve tens of thousands of stellar-mass binary systems at mHz frequencies.
The vast majority of these will be double white dwarfs in our Galaxy. 
The greatest part will remain unresolved, forming an incoherent stochastic foreground signal.
Using state-of-the-art Galactic models for the formation and evolution of binary white dwarfs and accurate LISA simulated signals, we introduce a test for foreground Gaussianity and stationarity, building on methods available for ground-based detectors.
We explain the observed non-stationarity with a new analytical modulation induced by the LISA constellation motion and the intrinsic anisotropy of the source distribution.
By demodulating the foreground signal, we reveal a deviation from Gaussianity in the 2~--~10 mHz frequency band.
Our finding is crucial to design faithful data models: the proposed method serves as a diagnostic and estimation tool to flag and model deviations, respectively. 
Neglecting them would introduce systematic biases on
individual sources and astrophysical foregrounds parameter estimation, 
ultimately leading to inaccurate interpretation of the LISA data.
\PACS{
      {04.80.Nn, 95.55.Ym}{Gravitational wave detectors}   \and
      {95.85.Sz}{Gravitational wave astronomical observations}
     } %
} %

\section{\label{sec:intro} Introduction}
The Laser Interferometer Space Antenna (LISA) will observe gravitational waves (GWs) from a large variety of physical systems emitting in the frequency range $\SIrange{e-4}{e-1}{\hertz}$.
In particular, LISA will observe millions of double white dwarfs (DWDs) in our Galaxy, and resolve tens of thousands of them~(for a review see Ref.~\cite{2023LRR....26....2A}). 
The superposition of their unresolved signals (quasi-monochromatic, though either too faint or too similar to be extracted individually from the data) will form a stochastic foreground --- a dominant source of noise at frequencies $\ssim\SIrange{0.1}{3}{\milli\hertz}$ --- often referred to as ``confusion noise''~\cite{2001astro.ph..8028P,2009CQGra..26i4030N}. 
Additional populations of astrophysical sources are also expected to form confusion backgrounds, e.g. from extra-Galactic DWDs~\cite{2024A&A...683A.139S}, Milky way satellites~\cite{2024MNRAS.531.2642R,2025PhRvD.111f3005P}, extreme-mass-ratio inspirals~\cite{2025PhRvD.111j3047P,2024CQGra..41k5004O,2023PhRvD.108j3039P,2024ApJ...965..148X,2023ApJ...955L..27N,2020PhRvD.102j3023B} and stellar-mass binary black holes~\cite{2023JCAP...08..034B}.

The incoherent nature of the confusion noise precludes its analysis as a deterministic signal.
However, it is possible to model, hence infer upon, its underlying statistical properties.
A large variety of confusion-noise spectral signatures have been modeled in the literature, focusing on the underlying population distribution and individual binary dynamics~\cite{Geo23,Kor22,2023MNRAS.525L..50S,2024A&A...692A.165T}.

Data-analysis pipelines for the estimation of LISA foreground signals have been developed under a number of simplifying assumptions: 
the majority of them assumes stationarity and Gaussianity, therefore uniquely describing foregrounds by their power spectral density (PSD)~\cite{2019JCAP...11..017C,2020CQGra..37u5017K,2020JCAP...07..021P,2021JCAP...01..059F,2023JCAP...04..066B,2024PhRvD.109d2001M,2021PhRvD.103j3529B,2023PhRvD.107l3531H}.
This is part of the larger effort to build a coherent, all-encompassing, data-analysis scheme for LISA data, including the estimation of individual sources, astrophysical foregrounds, and instrumental noise parameters~\cite{2025PhRvD.111b4060K,2024PhRvD.110b4005S,2023PhRvD.107f3004L,2024PhRvD.110j4069K,2025PhRvD.111j3014D}.
Alternative approaches have been proposed to devise parametrized models for long-term periodic non-stationarities, frequently referred to as ``cyclostationarities''~\cite{2022ApJ...940...10D,2025JCAP...04..052H,2024JCAP...06..055M,2006PhRvD..73l2001T,2005PhRvD..71l2003E}, or to directly model source distribution anisotropies~\cite{2025PhRvD.111b3025C,2021MNRAS.507.5451B}.
Similarly, recent progress has been made on the development of heavy-tailed likelihoods to infer statistical properties of non-Gaussian signals~\cite{2023PhRvD.108j3005S}.
The Galactic spatial distribution, together with the peculiar source frequency distribution in the LISA band, makes the confusion noise a prime candidate for both extensions. 
In this work, we employ a realistic model of the Milky Way DWD population, and connect it directly to the LISA datastream. 
We focus on spatial and frequency distributions as the main drivers of foreground Gaussianity and stationarity~\cite{2020PhRvD.102d3502C,2024PhRvD.109h3029P,2023PhRvD.108j3005S,PhysRevD.104.043019}.
In doing so, we provide a diagnostic tool to reveal such features, and the required formalism to explain them directly with the underlying astrophysical population.

In addition, as pointed out in Ref.~\cite{2024arXiv241017180R}, residual non-Gaussianities may arise in the residual data of a Global Fit after identification and conditioning on resolvable sources: we argue that our proposed test statistics is suitable for rapid residual consistency check.

The paper is organized as follows.
In Sec.~\ref{sec:astro}, we provide a concise overview of our approach to construct a representative synthetic population of Galactic DWDs.
In Sec.~\ref{sec:data}, we introduce our model for the LISA datastream and provide some key definitions.
In Sec.~\ref{sec:foreground}, we conduct a preliminary analysis of the stochastic foreground signal. 
We characterise the source parameter distributions and signal cross-contamination as probes to identify frequencies where target deviations are expected.
In Sec.~\ref{sec:stats}, we detail a description for the stationarity and Gaussianity of stochastic timeseries.
We then introduce a frequentist test sensitive to both, inspired by methods developed for ground-based detectors~\cite{2016PhRvD..93h2005Y,2023CQGra..40r5006A,2021CQGra..38m5014D}.
In Sec.~\ref{sec:envel}, we describe a semi-analytical model for the foreground signal envelope. 
A detailed derivation is provided in~\ref{app:envelopes}.
In Sec.~\ref{sec:results}, we present the results of our test applied to simulated LISA signals.
Finally, in Sec.~\ref{sec:discussion} we discuss the implications of our findings and the potential impact on the analysis of LISA data.

\section{\label{sec:astro} Galactic DWD population}

In this section, we briefly summarize the methodology used to assemble a representative synthetic population of Galactic DWDs detectable by LISA. 
This is a two-step process: 
first, we adopt a model to describe the DWD evolution 
and, second, we integrate this with a representation of the Milky Way.
 
We employ the DWD evolution models of Ref.~\cite{Too12}, 
which are generated using the rapid binary population synthesis code \texttt{SeBa}~\cite{SeBa,2001A&A...375..890N,Too12}. 
In these models, binaries are evolved starting from the Zero-Age Main Sequence until DWD formation. 
We stress that these synthetic models are calibrated for DWDs against available observations, 
matching both the mass-ratio distribution and the local number density~\cite{Too12,Too17}. 
From the \texttt{SeBa} models, we extract DWD binary component masses, 
orbital separations at DWD formation, and DWD formation time.

Next, we consider the stellar density distribution and star formation history of the Milky Way. 
These choices influence individual DWD distances and sky positions, 
as well as their present-day orbital and GW frequencies. 
The stellar density distribution of our model is made of two components: a central bulge and an extended stellar disk. 
The former is described by an exponential radial stellar disk profile with an isothermal height distribution, 
while the former follows a spherically-symmetric exponential distribution (for details, see~\cite{2004MNRAS.349..181N,Kor19}). 
These choices, as well as the scale parameters describing the density profiles, are motivated by observations~\cite{BH16}. We limit this study to a two-component (bulge + disc) model as the stellar halo contributes negligibly to the overall Galactic GW signal~\cite{2009ApJ...693..383R}. 

To describe the star formation history of the Galaxy, we use the plane-projected star formation rate from a chemo-spectrophotometric model by~\cite{BP99} for the stellar disc while doubling the star formation rate in the inner \SI{3}{\kilo\parsec} for the bulge. We post-process the orbital parameters of the binaries by accounting for gravitational-wave emission from DWD formation until the present age of the Galaxy, here assumed to be \SI{13.5}{\giga\year}.
As a result, our Galaxy formed inside-out, such that the median age of binaries is approximately \SI{10}{\giga\year} in the bulge, decreasing to around \SI{3}{\giga\year} at the Solar radius (\SI{8.2}{\kilo\parsec}, e.g.,~\cite{GRAVITY19}), and further decreasing to about \SI{2}{\giga\year} at the outskirts of the disk, as expected from observations (e.g., in~\cite{mak17}). 
With the assumptions above, we obtain a total stellar mass of \SI{8.2e10}{\solarmass}.

As a result of our modeling, we find approximately \num{26e6} DWDs emitting GWs in the LISA band. 
Each binary is described by a set of 8 parameters $\theta$, namely the GW initial frequency $f_0$, its derivative $\dot{f}$, amplitude $A$, ecliptic latitude $b$, ecliptic longitude $l$, inclination $\iota$, initial phase $\phi_0$, and polarization angle $\psi$.
$f_0$ and $\phi_0$ are defined as measured at the start of the LISA mission.
The first five parameters represent the outcome of the modeling procedure described above, while the remaining three angular parameters are assigned randomly: $\iota$ is sampled from a uniform distribution in $\cos \iota$, while $\psi$ and $\phi_0$ are sampled from a uniform distribution in $[0, 2\pi]$.
Our model assumes that each binary evolves in isolation and that interactions between its individual component objects (e.g., via tides) are negligible from DWD formation until present time (for a more general phenomenology, see e.g.~\cite{2023LRR....26....2A} and references therein). 
Thus, the frequency derivative is completely determined by the GW-driven orbital decay 
\begin{equation}
    \dot{f} = \frac{96 \pi^{8/3}}{5 c^5}  \left( G {\cal M}_c \right)^{5/3} f^{11/3}, \label{eq:fdot}
\end{equation}
where ${\cal M}_c$ denotes the source chirp mass 
\begin{equation} 
{\cal M}_c=(m_1m_2)^{3/5}/(m_1+m_2)^{1/5} ,
\end{equation} 
with $m_1>m_2$ being the binary component masses. 
The amplitude of the GW signal is given by
\begin{equation}
    A = \frac{2\left(G {\cal M}_c\right)^{5/3} \left(\pi f \right)^{2/3}}{c^4 D} ,\label{eq:amplitude}
\end{equation}
where $D$ is the distance, and $G$ and $c$ are the gravitational constant and the speed of light, respectively.

Only about \SI{1}{\percent} of the entire Galactic DWD population~---~mainly those with frequencies $f>\SIrange{2}{3}{\milli\hertz}$ or $f<\SIrange{2}{3}{\milli\hertz}$ but located nearby~---~are expected to be individually detectable by LISA with signal-to-noise ratio (SNR) $\gtrsim 7 $~\cite{2023LRR....26....2A}. 
The rest of the population will contribute to an unresolved stochastic foreground that we aim to characterize.  

\section{\label{sec:data} LISA Data processing}

LISA data are collected in the form of six raw interferometric outputs, each associated with a one-way laser link between two spacecrafts.
Individual link signals are then linearly combined after suitable delays to form Time-Delay Interferometry (TDI) variables, which suppress the laser phase noise~\cite{2021LRR....24....1T}.
In Sec.~\ref{sec:foreground}, we consider first-generation TDI variables, and denote them as $X$, $Y$, and $Z$.
To simulate single-link observables, we apply to each GW strain signal a time-dependent response.
In addition, we apply a linear transformation to the $(X,Y,Z)$ vector to diagonalize the noise covariance matrix.
We denote the resulting set of TDI variables as $A$, $E$, and $T$, or collectively as $h=(A,E,T)$. 

The LISA response effectively induces a periodic Doppler and amplitude modulation on each signal, which can be modeled as a function of the source ecliptic latitude, longitude, 
inclination and polarization, together with the initial position and orbital motion of each satellite.
We summarize the response properties relevant to this work in Sec.~\ref{sec:envel}.
In~\ref{app:envelopes}, we provide a detailed derivation followed by an analytical model for the resulting foreground-signal envelope, arising from the incoherent superposition of a population of signals. 

We simulate TDIs sampled with a cadence of \SI{15}{\second}, the Nyquist frequency being well above the highest GW frequency content in the foreground, $f\lesssim\SI{5}{\milli\hertz}$.
By construction, instrumental noises in the three TDIs are Gaussian and statistically uncorrelated, hence we model them through three PSDs, $S_{n,A}(f)$, $S_{n,E}(f)$, and $S_{n,T}(f)$~\cite{LISA2018}.
Once a signal is simulated, we evaluate its SNR~---~a measure of its detectability relative to the expected noise~--- as follows
\begin{equation}
    {\rm SNR} = \sqrt{\sum _k (h_k|h_k)_k}.
    \label{eq:snr_tdi}
\end{equation}
In Eq.~\eqref{eq:snr_tdi}, $(a|b)_k$ denotes the inner product of two TDI datastreams $a$ and $b$, weighted by the corresponding $S_k(f)$ which reads
\begin{equation}
    (a|b)_k = 4{\rm Re}  \int_{0}^{\infty} {\rm d}f  \, \frac{\widetilde{a}^*(f) \widetilde{b}(f)}{S_{n,k}(f)},
\label{eq:inner_prod}
\end{equation}
and summation is performed over elements of $h$, i.e. $A,E,T$.
A source is classified as ``detectable'' if its signal SNR exceeds a threshold of 7 and ``undetectable'' otherwise.
We perform such classification using the \textsc{Balrog} code, a suite of tools used to perform a variety of inferences on LISA sources and instrumental artifacts.
This includes supermassive binary black
hole mergers~\cite{2023PhRvD.108l4045P}, double WDs and NSs~\cite{2019PhRvD.100h4041B,2020ApJ...894L..15R,2023MNRAS.522.5358F,2024MNRAS.531.2817M}, stellar-mass binary
black holes~\cite{2021PhRvD.104d4065B,2022arXiv220403423K,2023PhRvD.108h4014B,2024PhRvD.110j3026B}, glitches~\cite{2023PhRvD.108l3029S}, and GW stochastic backgrounds~\cite{2024PhRvD.109h3029P}.

In a realistic scenario with numerous sources, the simultaneous presence of DWD signals affects their individual detectability.
We quantify the similarity between signals (hence their cross-contamination) by their mutual overlap
\begin{equation}
    \mathcal{O}_{ij} = \frac{\langle h^{(i)}\mid h^{(j)}\rangle}{\sqrt{\langle h^{(i)}\mid h^{(i)}\rangle\langle h^{(j)} \mid h^{(j)}\rangle}}, \label{eq:overlap}
\end{equation}
where the indexes $i$ and $j$ label the DWDs in the simulated catalog and 
\begin{equation}
    \langle a \mid b \rangle = \sum_k \left(a_k\mid b_k\right)_k.\label{eq:full_inner_prod}
\end{equation}
The overlap matrix ${\cal O}$ quantifies the correlation between signals and thus our capacity to resolve them individually. 
This is readily shown by inspection of the likelihood used to infer on a single-source model with parameter $\theta$
\begin{equation}
    \log {\cal L}(d\mid \theta) = -\frac{1}{2}\left\langle d - h(\theta)\mid d - h(\theta) \right\rangle + {\rm const.}\label{eq:single-like} 
\end{equation}
when the data contain multiple signals with parameters $\lbrace \theta_k\rbrace_{k=1}^{S}$, i.e..
\begin{equation}
d=n+\sum_{k=1}^S h(\theta_k).
\end{equation}
In fact, if the inferred parameter exactly matches the true value $\theta_j$ of one source in the data, the (much simpler) log-likelihood value one would obtain on single-source data $d_j=n+h(\theta_j)$ is corrected by additional terms
\begin{align}
\log {\cal L}(d\mid \theta) -\log {\cal L}(d_j\mid \theta) &\approx - \sum_{i\neq j} \langle h(\theta_i) \mid h(\theta_j)\rangle \label{eq:like-multi}
\end{align}
Therefore, when the overlap between two given sources is close to unity and the two sources have comparable SNRs, the additional likelihood terms in Eq.~\eqref{eq:like-multi} read 
\begin{align}
{\cal O}_{ij} {\rm SNR}_i{\rm SNR}_j \approx {\rm SNR}_i^2 \approx {\rm SNR}_j^2  \label{eq:overlap-contamination}
\end{align}
yielding false peaks in the full likelihood
, i.e. source confusion. 

This is the reason why ${\cal O}_{ij}$ is often used as an approximate metric to quantify inference contamination due to multiple source signals being present in the data~\cite{2004PhRvD..70h2004C}.
In Sec.~\ref{sec:foreground} we show the expected overlap between DWDs, describe it in terms of the source parameter distribution, and discuss the robustness of our results in light of the frequencies where it becomes dominant.
\begin{figure}[t]
    \centering
    \includegraphics[width=1.0\columnwidth]{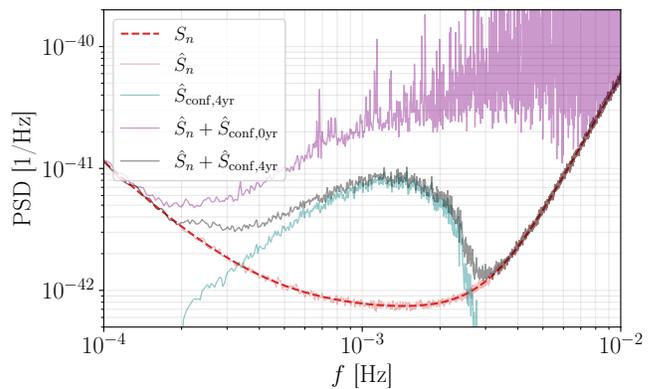}
    \caption{PSDs of dominant noise sources in LISA.
    The solid and dashed red curves denote the PSD analytical model and an estimate from a realization of the instrumental noise, respectively.
    The purple curve denotes the confusion noise PSD expected at the start of the mission. This is simulated using the full DWD population of Sec.~\ref{sec:astro}, assuming no source is detected, yet.
    After a nominal observation of $\SI{4}{\year}$, the residual confusion noise PSD after identification of resolvable sources is shown as a teal curve. 
    Its sum with the instrumental noise is shown as a solid black curve.}\label{fig:11}
\end{figure}

In the LISA context, the sum in Eq.~\eqref{eq:like-multi} is large. As mentioned in Sec.~\ref{sec:astro}, as many as $\num{26e6}$ DWDs are expected to be unresolvable; 
they build up to form a foreground, whose spectral density is expected to be larger than the instrumental noise at frequencies from \SIrange[range-phrase=~up to~]{0.1}{5}{\milli\hertz}.
Consequently, one needs to take it into account as an additional noise source with an overall amplitude that depends on the observation time: as more individual sources are resolved the effective noise level lowers.
Analytical spectral models incorporating such time dependence are available in literature~\cite{2021arXiv210801167B}, and we employ them to evaluate the SNRs in Sec.~\ref{sec:foreground}.
We denote analytical estimates of the instrumental noise spectral density as $S_n(f)$. 
Similarly, the confusion noise after an observation time $T$ is denoted by $S_{{\rm conf},T}$.
Instead, numerical estimates obtained applying the Welch algorithm~\cite{1967ITAE...15...70W} to individual realizations are denoted with $\hat{S}_n(f)$ and $\hat{S}_{{\rm conf}, T}$, respectively.
More recently, Ref.~\cite{PhysRevD.104.043019} devised an approximate method, often referred to as ``iterative foreground estimation'' (IFE), which goes beyond the individual source resolvability criterion based solely on the SNR. 

\begin{figure*}[t]
    \includegraphics[width=2.\columnwidth]{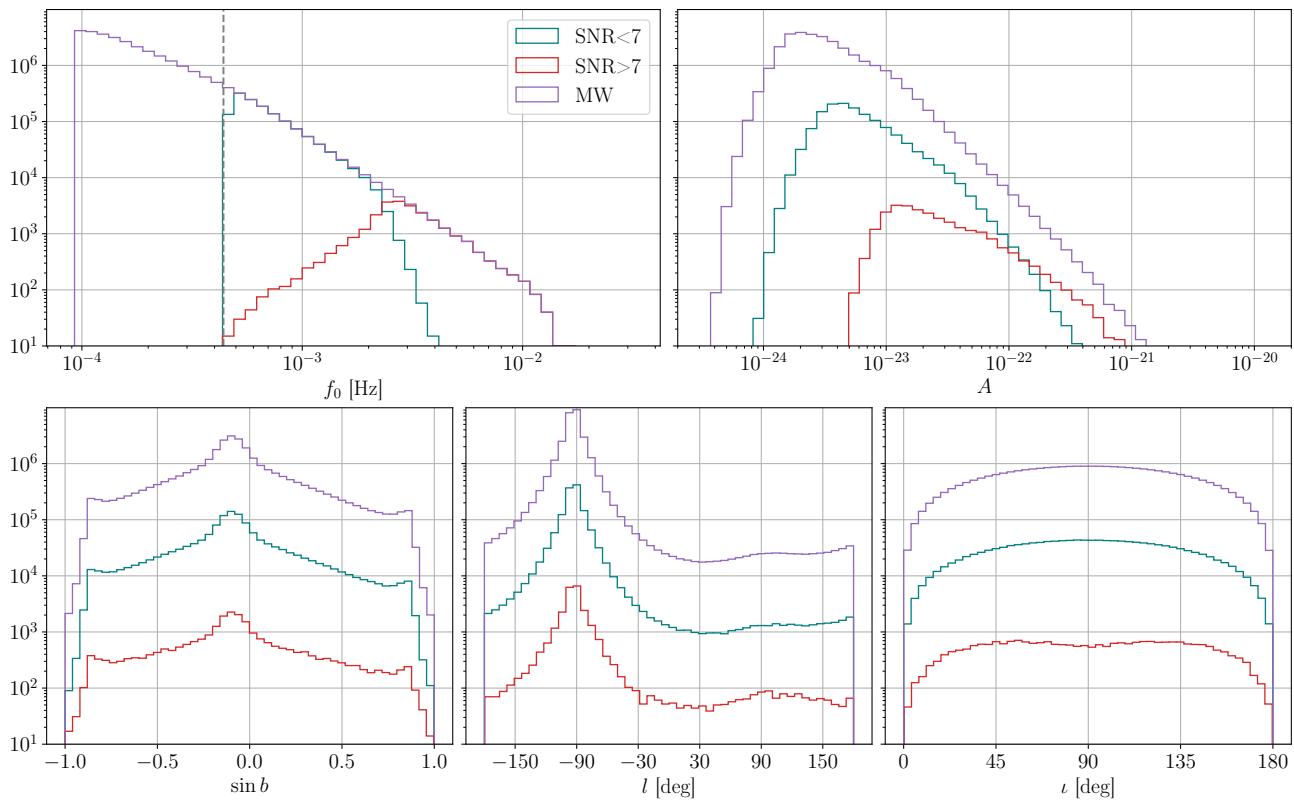}
    \caption{Population distribution of DWDs in the LISA band. The full population is shown in purple, containing approximately \num{2e7} sources.
    Teal (red) solid curves denote the undetectable (detectable) subpopulation.
    The top panels show the source distribution in frequency $f_0$ and amplitude $A$.
    The bottom panels show the source distribution in sine-ecliptic latitude $\sin b$, longitude $l$, inclination $\iota$.
    Our model is constructed following prescriptions described in Sec.~\ref{sec:astro}. 
    We simulate each source individually following the procedure described in Sec.~\ref{sec:data}, and classified it as resolvable if its SNR exceeds a threshold of $7$. 
    About $1\%$ of the whole population is classified as such, the largest majority above $\SI{1}{\milli\hertz}$ (red histograms).
    The reminder (teal histograms) remains unresolvable.
    Below \SI{0.44}{\milli\hertz} (shaded gray area), fewer than $1$ in $10^6$ sources are resolvable. 
    }\label{fig:1}
\end{figure*}

In Sec.~\ref{sec:results}, we use the classification obtained by the IFE algorithm to simulate again each signal with \textsc{Balrog} and provide a more faithful realization of the foreground.

In Fig.~\ref{fig:11} we show representative realizations of the confusion noise at both the start of the mission and after four years of observation, as purple and teal solid curves, respectively.
A realization of the instrumental noise and its underlying analytical model are shown as red solid and dashed curves, respectively.

\section{\label{sec:foreground} Galactic foreground properties}

We conduct a preliminary analysis of the foreground by generating individual signals and computing their resolvability in isolation (i.e. based solely on their SNR). 
In Fig.~\ref{fig:1}, we show the distribution of the DWD population in the LISA band, alongside the two subpopulations of detectable and undetectable sources.
We show both as a function of source GW initial frequency $f_0$, amplitude $A$, sine-ecliptic latitude $\sin b$, ecliptic longitude $l$, and inclination $\iota$.

The majority of the detectable sources have $f_0>\SI{0.4}{\milli\hertz}$.
This is due to the LISA sensitivity curve steeply increasing at lower frequencies and to a milder decrease of the GW amplitudes, $A\propto f_0^{2/3}$. 
Overall, sources with amplitude $A$ below $\num{e-23}$ are systematically undetected.
We additionally explore the detectability as a function of extrinsic parameters, relevant for the foreground modulation discussed in Sec.~\ref{sec:envel}.
We report a mildly higher detectability for sources high in the ecliptic plane, as shown by the narrow peaks around $\left|\sin b\right|\approx 0.9$.
Similarly, a slightly higher detectability is found for sources with inclinations close to face-on (face-off), i.e. $\cos \iota \approx 1$ ($\cos \iota \approx -1$), relative to edge-on cases with $\cos \iota \approx 0$.
The higher detectability for sources at $\left| l- 90^\circ \right|\lesssim 30^\circ$ depends significantly on the assumed initial position of the LISA constellation, which is yet to be established.

To improve the SNR-based classification, we additionally quantify the overlap between unresolvable sources.
Evaluating ${\cal O}$ from Eq.~\eqref{eq:overlap} over the whole population is computationally prohibitive, as it requires the generation and storage of $\approx \num{e7}$ signals and the evaluation $\approx\num{e14}$ overlaps.
We coarse-grain it over chunks of $N_c=\num{50}$ sources neighbouring each other in frequency, hence reducing the number of overlap evaluations to \num{5e8} and eliminating the need for storage.
In practice, we evaluate the average overlap 
\begin{equation}
    \overline{\cal O}_{i} = \left\langle {\cal O}_{(i+j)(i+k)} \right\rangle_{j,k=1,\ldots,N_c},
    \label{eq:av-overlap}
\end{equation}
the median frequency $\langle f_0 \rangle$, and the average frequency width $\Delta_{f_0}$ over each chunk.
The latter is the sum of the Doppler modulation and the intrinsic, narrower frequency drift due to the GW-driven orbital tightening in Eq.~\eqref{eq:fdot}.
While the former depends on the sky-position of each source, the latter does not depend on extrinsic parameters and is influenced by the mission duration, only.  

\begin{figure}[t]
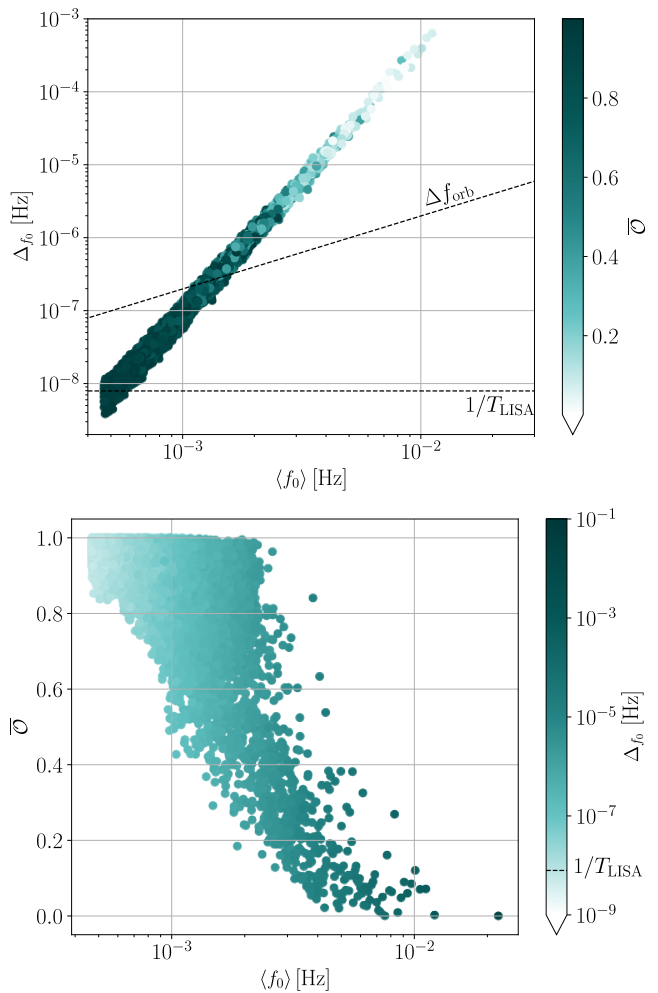

    \centering
    \includegraphics[width=1.0\columnwidth]{Figure3top.pdf}
    \includegraphics[width=1.0\columnwidth]{Figure3bottom.pdf}
    \caption{
    Correlation and frequency drift of unresolvable DWDs in LISA.
    We show the average overlap as defined in Eqs.~\eqref{eq:overlap} and \eqref{eq:av-overlap} and the average frequency width of individual signals, as a function of the average frequency over chunks of 50 sources, neighbouring in frequency.
    Relevant frequency scalings are discussed in Sec.~\ref{sec:foreground}.
    }\label{fig:6}
\end{figure}

Figure~\ref{fig:6} (top panel) shows a trend in $\Delta_{f_0}$ close to \phantom{    }
$\num{e-7} (\langle f_0 \rangle/\SI{1}{\milli\hertz})^{3.6}$,
which maps in Fig.~\ref{fig:6} (bottom panel) to an average overlap that scales roughly as $\propto \langle f_0 \rangle^{-1.0}$ (though with larger variability) across sources.
Two frequency scales are relevant in both figures:
first, the intrinsic LISA frequency resolution after $T=\SI{4}{\year}$ of observation and, second, the maximal frequency drift induced on a source on the ecliptic plane, $b=0$.
Source non-resolvability below \SI{0.5}{\milli\hertz} is not a consequence of the former, alone. 
In fact, LISA orbits will be known by telemetry, hence incorporated in the time-dependent response as part of the likelihood model.
This effect becomes negligible when the Doppler modulation itself is small, i.e. for sources close to the ecliptic poles. 
However, less than $5\%$ fall in the range $\left| b\right|> \SI{45}{\degree}$.
Overall, as we will see in Sec.~\ref{sec:results} and illustrate in Fig.~\ref{fig:8}, the strongest deviation from Gaussianity occurs at and above $\approx \SI{3}{\milli\hertz}$ where the overlap is only mild, ${\cal O} \lesssim 0.4$.
Therefore, results obtained using the IFE algorithm are robust: assessing the detectability solely based on individual source SNRs and neglecting the correlations introduced in Eq.~\eqref{eq:overlap} and Eq.~\eqref{eq:overlap-contamination}, is a good proxy for the residuals of a perfect global fit, i.e. one not introducing additional non-Gaussianities and non-stationarities from, e.g., modelling mismatches.

\section{\label{sec:stats} Statistical framework}

We detail below a description of stationary and Gaussian stochastic timeseries, leading to the formulation of a statistical test sensitive to both properties.

\subsection{\label{sec:process} Stationary Gaussian processes}
A foreground can be modeled by a probability distribution over functions, e.g. $x(t)$, or in its discretized form by a multivariate distribution over their samples $\left\lbrace x(t_i)\right\rbrace_{i}$.
Gaussian processes have the exclusive peculiarity of being completely specified by a finite number of correlators, i.e.~their mean and covariance functions~\cite{Isserlis}.
In the time domain, this is summarized by the following properties
\begin{align}
    x(t) &\sim {\cal N}(\mu(t), \Sigma(t,t^\prime)) , \label{eq:prob-td}\\
    \mu(t) &= \langle x(t)\rangle ,  \label{eq:mean-td}\\
    \Sigma(t,t^\prime) &= \langle x(t) x(t^\prime)\rangle , \label{eq:cov-td}
\end{align}
where angled brackets denote average over ensemble, $\mu(t)$ and $\Sigma(t,t^\prime)$ denote mean and covariance function, respectively.

To overcome the availability of only a single astrophysical realization of the process, ergodicity is often assumed. 
If the observation time is larger than the inverse minimum non-zero frequency of interest ---~ as it is the case for LISA, with $T_{\rm LISA} = \SI{4}{\year}$ and $f_{\rm low}>\SI{2e-5}{\hertz}$~--- averages over ensemble are estimated trading off frequency resolution for the availability of multiple signal realizations. 
This is, e.g., the same procedure followed by the Welch algorithm to obtain a PSD estimator through multiple periodograms from (possibly overlapping) data segments. 
In LISA, this amounts to approximately \num{e3} non-overlapping realizations of \num{e5} seconds each.

If the process is also \emph{second-order weakly stationary}~\cite{2020arXiv200910316M,1992PhRvD..46.5236F}, the covariance in Eq.~\eqref{eq:cov-td} effectively depends only on the difference $t-t^\prime$.
In Fourier domain, this is equivalent to 
\begin{equation}
 \label{eq:stationarity-fd}  
 \langle {\tilde x}^*(f){\tilde x} (f^\prime) \rangle \propto \delta(f-f^\prime) S_x({f}),
 \end{equation}
where $\delta(f-f^\prime)$ 
denotes a Dirac delta in frequency domain, ${\tilde x}(f)$ is the Fourier transform of $x(t)$
\begin{equation}
    \tilde x(f) = \int^\infty_{-\infty}{\rm d}t\ x(t) e^{-{\rm i}2\pi ft} ,
    \label{eq:FD}
\end{equation}
and $S_x({f})$ is the process PSD, related to the covariance function~\cite{khintchine1934korrelationstheorie} 
\begin{equation}
    S_x(f)= \int^\infty_{-\infty}{\rm d}t\ \Sigma (t,t^\prime) e^{-{\rm i}2\pi f (t-t^\prime)}. \label{eq:psd-from-twopoint}
\end{equation}

\subsection{\label{sec:rail} A frequentist, frequency--domain test}

The null-hypothesis we want to test is that satisfied by an ergodic, zero-mean, second-order weakly stationary Gaussian process, i.e. completely specified by its spectrum through Eq.~\eqref{eq:psd-from-twopoint}.

The Fourier transform in Eq.~\eqref{eq:FD} is a linear operator, which implies the process is equivalently described by two independent Gaussian variables
\begin{equation}
    \Re \tilde{x}(f), \Im \tilde{x}(f) \sim {\cal N}(0, S_x(f)),\label{eq:realpart}
\end{equation}
where $\Re$ ($\Im$) denotes the real (imaginary) part of a complex number, respectively.
The norm of the complex variable $\tilde{x}(f)$ is therefore distributed as an infinite sequence of Rayleigh variables
\begin{equation}
    \left| \tilde{x}(f) \right| \!=\! \sqrt{\Re \tilde{x}(f)^2 + \Im \tilde{x}(f)^2} \sim {\rm Rayleigh}(S_x(f)).\label{eq:rayleighdistro}
\end{equation}
Its square is distributed as 
\begin{equation}
    \left| \tilde{x}(f) \right|^2 \sim \Gamma(1,2S_x(f)),\label{eq:gammadistro}
\end{equation}
where $\Gamma(k,\theta)$ denotes the Gamma distribution and $k,\theta$ are the defining shape and scale parameters. 
The Gamma distribution mean and variance are given by
\begin{align}
    \langle y \rangle &= k \theta, \label{eq:gamma-mean}\\
    \langle (y- \langle y \rangle)^2\rangle &= k \theta^2 \label{eq:gamma-variance}
\end{align}
respectively. 
Therefore, the \emph{coefficient of variation}, defined as the ratio between the square-root of the variance in Eq.~\eqref{eq:gamma-variance} and the mean in Eq.~\eqref{eq:gamma-mean}, serves as a test statistic $\rho_{[x]}$, whose value for stationary Gaussian processes is frequency-independent and identically one across its whole domain, i.e.
\begin{equation}
    \rho_{\left[x\right]}(f) =\frac{\sqrt{\langle (y - \langle y \rangle )^2\rangle }}{\langle y \rangle } = 1.
    \label{eq:nullhyp}    
\end{equation}
In Eq.~\eqref{eq:nullhyp}, the quantity $\rho_{\left[ x \right]}(f)$ denotes the operator mapping a realization of the process $x$ onto its test statistic as a function of frequency.

Processes whose Fourier-transform squared norm fluctuates around their mean more than the mean itself yield a test statistic larger than one. This is the case, e.g., of cyclostationary Gaussian processes: they are defined similarly to stationary processes, where the two-point correlation function $\Sigma(t,t^\prime)$ exhibit a short-term dependence on the difference $t-t^\prime$ and a periodic long-term one on, e.g., $t+t^\prime$ with period $T$.
In the frequency domain, the correlation function can be conveniently decomposed using a Fourier-series
\begin{equation}
    \Sigma(t,t^\prime) = \sum_{n=-\infty}^{\infty} \Sigma_{n}(t-t^\prime) e^{-2\pi {\rm i} n t/T},
\end{equation}
where $\Sigma_{n}(t-t^\prime)$ are the Fourier coefficients of the covariance function
\begin{equation}
    \Sigma_{n}(\tau) = \frac{1}{T}\int_{0}^{T} {\rm d}t\ \Sigma(t,t +\tau) e^{-2\pi {\rm i} n t/T},
\end{equation}
and their Fourier transform are often referred to as \emph{cyclic spectra}. 
As $T$ approaches infinity, the process becomes stationary and the Fourier coefficients $\Sigma_{n}$ are suppressed except for the covariance function $\Sigma_0(\tau)$.
Such processes are not \emph{auto-covariance ergodic} over timescales comparable to $T$: the covariance estimated from segments of a realization has contributions from $\Sigma_0(\tau)$ and from higher-order terms $\Sigma_{n>0}$, while they preserve the same mean PSD.
The opposite limiting case is that of a deterministic signal $h(t)$, interpreted as a realization $x(t)$ of a stochastic process. 
Its Fourier transforms ${\tilde x}(f)$ are distributed at each frequency as $\delta(x(f)-h(f))$, yielding effectively zero variance, hence $\rho_{[h]}(f)=0$.
We anticipate a similar scenario in our study, at frequencies between \SIrange[range-phrase=~and~]{3}{6}{\milli\hertz} where the foreground is dominated by just a few unresolved sources, see Fig.~\ref{fig:1}. 

In practice, the test is carried out constructing estimators for each random quantity in Eq.~\eqref{eq:nullhyp}.
The denominator is evaluated via the Welch's PSD estimator~\cite{1967ITAE...15...70W} while the numerator is obtained through individual FFTs.

\begin{figure}[t]
    \centering
    \includegraphics[width=1.0\columnwidth]{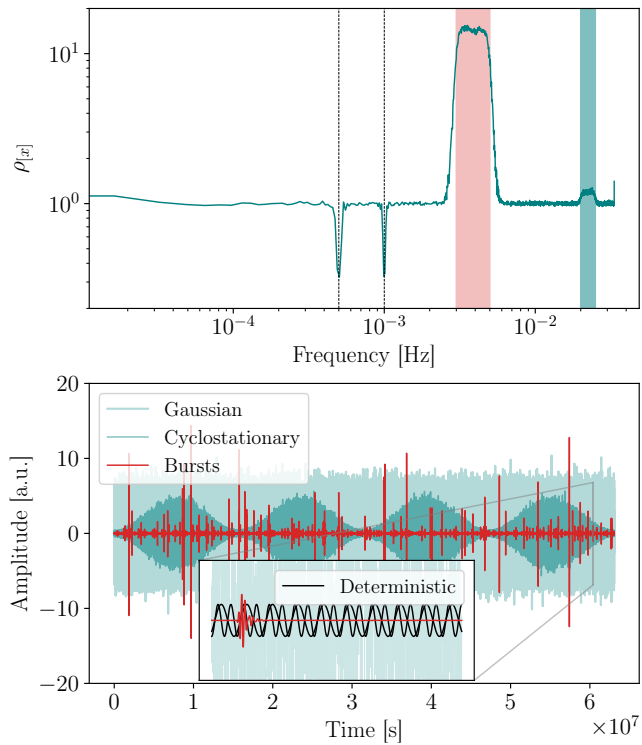}
    \caption{Toy model showing the response of our test to violations of stationarity and Gaussianity (teal solid line, top panel). 
    At frequencies of $\num{0.5}$ and $\SI{1}{\milli\hertz}$ (black dashed lines, top panel) violations of Gaussianities appear due to the two monochromatic, deterministic signals injected in the data (black solid curve, bottom panel inset).
    In the range $\SIrange{3}{5}{\milli\hertz}$ (red shaded area, top panel) the test reveals violations of stationarity due to the simulated, suitably bandpassed, heavy-tailed Cauchy noise (red solid lines, bottom panel and inset). This is a proxy for a superposition of transient unresolved signals.
    For completeness, we show a $\SI{5.4}{\hour}$ segment of the data to reveal the structure of deterministic and burst signals.  
    In the range $\SIrange{20}{25}{\milli\hertz}$ (teal shaded area, top panel) the test exhibits violations of stationarity due to the Gaussian, bandpassed noise modulated by a long-term trend with period $T=\SI{1.5e7}{\second}$ (dark teal solid curve in bottom panel). 
    A white broadband Gaussian noise is also injected across the whole frequency range (light teal solid curves in bottom panel), yielding no test statistic violation. 
    The effect of different violations on the test statistic is discussed in Sec.~\ref{sec:stats}. Their significance is established in Sec.~\ref{sec:results}, where we apply the test on simulated LISA data.\label{fig:toy}}
\end{figure}

In Fig.~\ref{fig:toy}, we show a toy model illustrating different test violations of stationarity and Gaussianity and how they are identified by the proposed test.
The signal in our toy-model is a sum of two sinusoids at $f=\SIrange{0.5}{1}{\milli\hertz}$, a Gaussian stationary noise across the whole band, a cyclostationary Gaussian noise with a period of $T=\SI{1}{\year}$ band-passed at frequencies in $\SIrange{20}{25}{\milli\hertz}$, and a Cauchy noise band-passed at frequencies in $\SIrange{3}{5}{\milli\hertz}$.
In order to obtain the latter two, we band pass broadband realizations with a fourth order Butterworth filter~\cite{Butterworth1930}.
The clear frequency separation of the injected signals let us illustrate intuitively the test behaviour in different regimes. 
On one hand, the two fully deterministic sinusoids (black curves in bottom panel, inset plot) have almost constant squared Fourier-amplitudes if sufficiently many oscillations are accumulated in each chunk.  
The arbitrarily small variances in Eq.~\ref{eq:nullhyp} yield $\rho[x] \ll 1$, as we observe at the injected frequencies (top panel).
On the other hand, the superposition of burst-like signals (red curves in bottom panel) produces a heavy-tailed distribution for $y$, alternating between segments of negligible power and bursts of high energy. The resulting distribution has larger variance as compared to that of a Gaussian, white, stationary signal. Deviations in $\rho[x]\gg 1$ arise at the associated frequency (red shaded area in top panel, at $f\approx \SI{4}{\milli\hertz}$).
An intermediate behaviour is observed for the cyclostationary signal (dark teal curve in bottom panel) and the test statistic deviation (teal shaded area in top panel) significance is controlled by its amplitude relatively to the superimposed white, stationary noise.

Critical values for the test statistic --- applied on a timeseries of finite duration and split in $N_c$ segments --- can be obtained under the null-hypothesis, i.e. for a perfectly Gaussian stationary signal of the same (finite) duration of the GWB datastream.
Asymptotically, for a finite number $N_c$ of segments, the test is distributed as a ${\cal N}\big(1, 1/2\sqrt N\big)$. 
We will use such scaling to establish the significance of our results in Sec.~\ref{sec:results}.
In order to further decouple long-term cyclostationarities from narrowband non-Gaussianities, in Sec.~\ref{sec:envel} we construct a semi-analytical model of the foreground envelope. 
This serves as a tool to further demodulate any long-term trend and robustly confirm the presence of non-Gaussianities.

\section{\label{sec:envel} Foreground envelope}

In order to derive an approximation of the envelope in time of the confusion foreground, we make a few helpful simplifying assumptions. 
At the low frequencies, in which the galactic binary signal exists, the LISA signal can be approximated by two independent effective Michelson interferometers corresponding to the A and E channels, referred to as the low frequency approximation. 
In this framework, we first derive the envelope of the LISA response as a function of time for a large number of sources located at one point in the sky by averaging the square signal over a period, and the inclination and polarization angles.
\begin{equation}
\left\langle h_{A,E} \right\rangle^2 (t, b_S, l_S)
= \frac{1}{T} \int_0^T  dt' \int_{-1}^1 d \cos \iota \int_0^{2\pi} d\psi \,  h_{A,E}^2.
\end{equation}

This results in envelopes as a function of time $t$, and the location of the sources in the sky $(b_S, l_S)$. The details of this derivation are provided in~\ref{app:individualenvelope}.
With this result in mind, we then model the projected 2D spatial distribution of unresolved binaries by a bivariate Gaussian 
\begin{align}
p(b_S, l_S) &= \frac{1}{2\pi \sigma_1 \sigma_2} \exp \left[ - \frac{1}{2} \bm{\alpha}^T \mathbb{R}^T \mathbb{M} \mathbb{R} \bm{\alpha} \right], \\
\bm{\alpha} &= \left( \begin{array}{c}
b_S - b_M \\
l_S - l_M \end{array} \right), \\
\mathbb{M} &= 
\left( \begin{array}{c c}
\sigma_1^{-2} & 0 \\
0 & \sigma_2^{-2} \end{array} \right), \\
\mathbb{R} &=
\left( \begin{array}{c c}
\cos \delta & -\sin \delta \\
\sin\delta & \cos \delta \end{array} \right),
\end{align}
centered at ecliptic coordinates $(b_M, l_M)$, with variances $\sigma_{1,2}$ along the principal axes, rotated by an angle $\delta$ with respect to the ecliptic. We use this model to derive the envelope of the LISA response to the unresolved galactic foreground as a function of time and the model parameters with
\begin{multline}
H_{A,E}^2 (t, b_M, l_M, \sigma_1, \sigma_2, \beta) \\
= \int_\mathbb{R} \cos b_S db_S \int_\mathbb{R} dl_S \ p(b_S, l_S) \left\langle h_{A,E} \right\rangle^2 (t, b_S, l_S).
\end{multline}

For simplicity, we perform the integral over the whole real line in the ecliptic parameters: this should not affect the results significantly provided that the model distribution falls off sufficiently quickly as the ecliptic parameters approach their physical boundary. We present the details of this derivation in~\ref{app:skydist}.

\begin{figure}[t]
    \centering
    \includegraphics[width=\columnwidth]{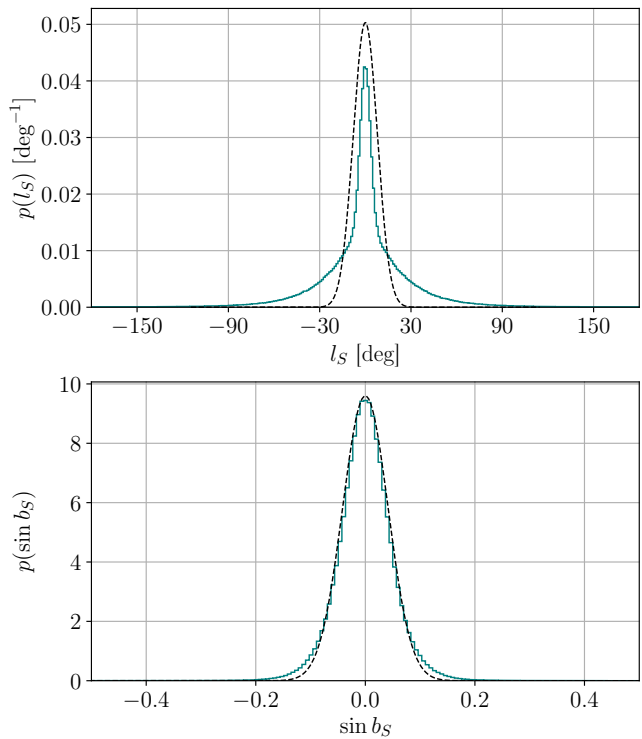}
    \caption{Spatial source distribution for our simulated MW population of DWDs presented in Sec.~\ref{sec:astro}. Sky-coordinates ($l_S, \sin b_S$) are shifted and rotated to yield zero mean and minimize their correlation. 
    While the latitude distribution is well approximated by a univariate Gaussian, the longitude--roughly corresponding to the Galactic longitude-- exhibits heavier tails.\label{fig:9}}
\end{figure}

\begin{figure}[t]
    \centering
    \includegraphics[width=1.0\columnwidth]{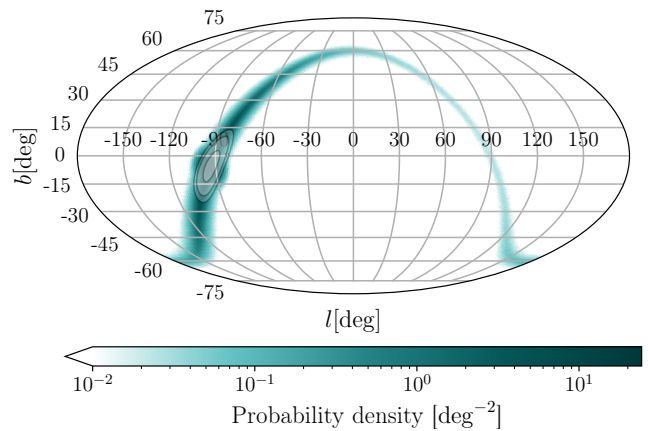}
    \caption{Distribution of sources in Ecliptic latitude and longitudes. Black ellipses and white nested shaded areas denote the $68\%$, $95\%$ and $99.7\%$ contour levels of the bivariate Gaussian approximation shown in Fig.~\ref{fig:9}.}\label{fig:10}
\end{figure}

We then fit the model parameters to the simulated galaxy presented in Sec.~\ref{sec:astro}. 
We show a comparison of our best-fit distribution with the simulated catalogue in Fig.~\ref{fig:9} and~\ref{fig:10}, and a comparison of the approximate envelope to a full simulation of the LISA data in Fig.~\ref{fig:8}. 
The simulated distribution is wider than the model, particularly along the ecliptic longitude. This is due to the Galaxy being not very well approximated by a single bivariate Gaussian distribution. 
Nevertheless, Fig.~\ref{fig:8} shows that our simple model is effective in representing the time dependence of the Galactic foreground. This could be improved by refining the model, e.g.\ by using a mixture of Gaussians or more realistic distributions. We leave this for future work.

\section{\label{sec:results} Results}

\begin{figure*}
    \centering
    \includegraphics[width=1.8\columnwidth]{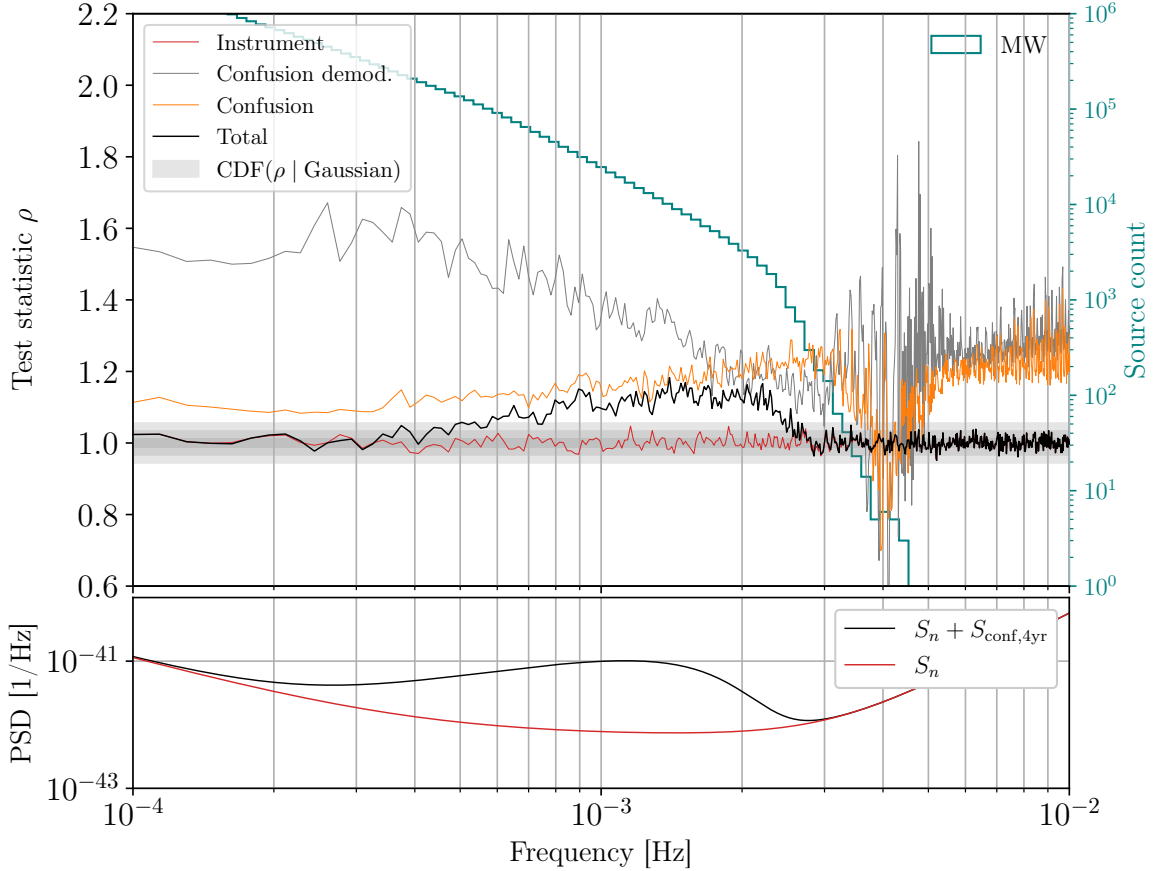}
    \caption{Our proposed test statistic applied to the LISA simulated foreground. We focus on the A channel, similar results holding for the E channel. 
    Critical values for the hypothesis test are shown as nested grey shaded areas at 68\%, 95\% and 99\% confidence.
    We test four timeseries and accompany them by the histogram of foreground sources count in frequency (teal solid histogram, top panel). 
    The test does not reveal violations for Gaussian instrumental noise (red solid curve, top panel). 
    The DWD foreground in isolation (orange solid curve), shows broadband violations of stationarity $(\rho_{[x]}>1)$ across all frequencies, and violations of Gaussianity $(\rho_{[x]}<1)$ around $\SI{4}{\milli\hertz}$.
    The latter arise from the foreground coherence between \SIrange[range-phrase=~and~]{3.8}{4.3}{\milli\hertz}, due to the limited number of sources contributing to it.
    The former arises instead from the coherent modulation of the foreground due to the LISA satellites' orbital motion.
    The test applied on the demodulated foreground (grey solid curve) shows similar deviations, and we discuss in Sec.~\ref{sec:results} a possible explanation.
    Finally, the test on the superposition of the instrumental noise and the foreground signal (black solid curve) yields violations of stationarity only at frequencies between $\SIrange[range-phrase=~{\rm and}~]{0.5}{3}{\milli\hertz}$. 
    The other violations are masked by the instrumental noise, whose PSD dominates (solid red line, bottom panel) over the foreground one (solid black curve, bottom panel) outside of the above interval.}\label{fig:12}
\end{figure*}

We now apply the proposed test on simulated LISA data.
Results are presented in Fig.~\ref{fig:12}, where four distinct timeseries are considered.
For brevity, we show results for the A channel, though we stress that the E channel behaves very similarly.
First, we cross-check our formalism against a realization of perfectly Gaussian instrumental noise. 
The test yields results largely compatible with the null-hypothesis across the whole frequency range, as shown by the critical values at 68\%, 95\% and 99\% credibility, shown as nested grey shaded areas.
Then, we test the DWD foreground in isolation as approximated by the IFE algorithm after \SI{4}{\year} of observation: 
broadband violations of stationarity are identified at all frequencies while violations of Gaussianity appear at frequencies around $\SI{4}{\milli\hertz}$.
The latter are due to individual contributions to the foreground from less than $40$ sources emitting between \SIrange[range-phrase=~and~]{3.8}{4.3}{\milli\hertz}, as revealed by the source count in frequency (top left panel in Fig.~\ref{fig:1}).
At such frequencies, the signals superposition does not suffice to build an effective incoherent signal, due to the limited number of sources contributing to it.
The former arise instead from the coherent modulation of the foreground due to the orbital motion of the LISA satellites.

\begin{figure}[th]
    \centering
    \includegraphics[width=1.0\columnwidth]{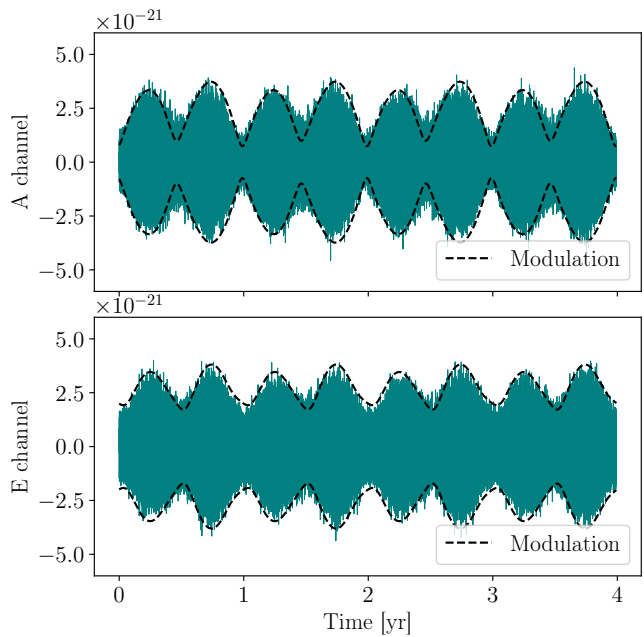}
    \caption{TDI time-domain envelope for a bivariate Gaussian source distribution over the sky. 
    The envelope reproduces globally the signal modulation over each period ($T=\SI{1}{\year}$), with limited inaccuracies close to the foreground maxima and minima. These are likely to arise from the rotated longitude ($l_S$) source distribution, which exhibits heavier-than-Gaussian tails, as shown in Fig.~\ref{fig:9}.}\label{fig:8}
\end{figure}

In order to confirm the origin of this non-stationarity, we demodulate in time-domain the foreground realization using the formalism developed in Sec.~\ref{sec:envel} as follows

\begin{equation}
    A_{\rm conf, demod.}(t) = \frac{A_{{\rm conf}, \SI{4}{\year}}(t)}{A_{\rm envelope}(t)},\label{eq:demodulation}
\end{equation}
and apply the test to the resulting signal.
We observe mainly two effects:
while the violation of stationarity is mildly reduced between \SIrange[range-phrase=~and~]{2.0}{3.0}{\milli\hertz}, it is amplified to a greater significance level outside this range.
We associate it to the limited accuracy of the envelope model close to each peak and valley of the signal, see Fig.~\ref{fig:8}.

Finally, we perform the test on the superposition of the instrumental noise and the foreground signal, yielding a somewhat surprising result:
non-Gaussianities are suppressed, while the non-stationarity persists at frequencies where the foreground spectrum dominates over the instrumental noise one.
This is due to the test being sensitive to violations in the amplitude of each segment Fourier transform.
Therefore the instrumental noise masks violations where its Fourier transform amplitude distribution dominates over that of the foreground.
This is further confirmed by the non-stationarity at the lowest frequencies of interest, disappearing below approximately $\SI{0.5}{\milli\hertz}$.
While the relative amplitudes of instrumental and confusion noise is expected to play a dominant role, additional factors contribute to the suppression or strengthening of the test statistic: e.g., the number of modulation cycles accumulated and the duration of the time-domain chunks relative to the modulation timescale. 
In addition, in a realistic LISA context, this is further complicated by the availability and length of uninterrupted data  segments. We leave a detailed, mathematically robust, study of such dependencies to future work.

\section{\label{sec:discussion} Implications for LISA analysis}

The results of our study suggest a number of implications for LISA data analysis. 
The test statistic in Eq.~\eqref{eq:nullhyp} is primarily suitable for application as a rapid diagnostic tool. 
Alternatively  ---~and similarly to the strategy proposed in Ref.~\cite{2024arXiv241017180R}~--- it can be used as a test on the residual data after identification and conditioning on resolvable sources. 
The distinct test response to deviations from Gaussianity or stationarity, exemplified in Fig.~\ref{fig:toy}, is an additional resource to consistently check the coherence of global fit results: should LISA data residuals yield significant deviations, the noise model employed in the Bayesian likelihood (often assumed Gaussian and stationary) needs to be extended, e.g. with methodologies similar to Ref.~\cite{2025PhRvD.111b2005K}.
Failure in doing so, may result in overestimation of the overall noise PSD to accomodate for amplitude drifts or fatter distribution tails. This in turn will effectively bias resolvable source parameter estimates, e.g. overestimating their luminosity distance. 
Flagging such data model ``failures'' in low-latency before informing EM counterpart follow-ups will be of central importance.  
We leave a detailed analysis of the impact on resolvable sources to future study.

By applying our test to representative LISA data we show that a non-stationary model for the Galactic confusion noise is required for realistic studies (as also pointed out in previous, time-frequency domain ones~\cite{2020PhRvD.102l4038C}).
Finally, a strong deviation of global fit residuals from stationarity may hint at the presence of yet unidentified backgrounds. 
In fact, recent studies have used the envelope presented in~\ref{app:envelopes} as a parameterized model to infer on putative additional backgrounds~\cite{2025PhRvD.111f3005P}, and distinguish them from the Galactic confusion noise.

A few additional details on the results presented are worth highlighting.
First, in this study we analyzed a synthetic population constructed based on specific modeling choices, some of which may impact our conclusions.
For instance, Ref.~\cite{Geo23} examined how the stochastic component changes when modifying the model of the Milky Way.
They found that altering the shape of the Galaxy does not noticeably affect the spectral shape of the stochastic foreground.
However, changing the total number of binaries in the LISA band, which effectively corresponds to changing the total stellar mass, has a notable impact.

For a fixed DWD binary evolution model, they demonstrated that increasing the number of binaries causes the overall confusion noise amplitude to increases, see also~\cite{2006ApJ...645..589B}.
This makes the resolvability of individual sources more challenging, leading to a milder reduction in stochasticity with frequency.
They also confirmed that a constant star-formation history, equivalent to our fiducial star-formation history over the past several~\si{\giga\year}, has no significant effect on the shape of the Galactic GW foreground.
Changing details in binary evolution assumptions may also lead to significant changes in both the number of resolved sources and the characteristics of the foreground.
For example, Ref.~\cite{Kor22} assembled an observationally-driven population of DWDs for LISA, employing the same Milky Way model as in this study.
However, their assumptions on the white-dwarf mass and frequency distributions at DWD formation rely on results obtained by interpreting DWD candidates in spectroscopic surveys~\cite{Mao12,Mao17,2018MNRAS.476.2584M}.
They found that, while the total number of DWDs in the LISA band is similar to that in our theory-driven model, the differences in the DWD properties result in a threefold increase in the number of individual detections, and changes in the shape of the unresolved foreground.
Notably, the latter extends to slightly lower frequencies, due to DWDs in the observationally driven model being, on average, easier to resolve by LISA.

Further modeling assumption of the foreground population and overall signal may impact our results:
while we employed the approximate result of the iterative foreground estimation algorithm, a more appropriate input would be a catalog of unresolved sources at various stages of a global-fit execution.
Similarly, a more flexible model for the source distribution in $l_S$ would increase the accuracy of our envelope model. We leave both for future work. 
However, we stress that the test statistic proposed is well-defined in and applicable to such scenarios. 
We also highlight that at the time of writing, independent studies have confirmed our findings following different methodologies~\cite{2025PhRvD.111b2005K,2024arXiv241017180R}.

Finally, we point out that, while our test is sensitive to the amplitude distribution of foreground segments in Fourier domain, the corresponding phases carry additional information: a stochastic Gaussian signal is expected to have phases uniformly distributed in $[0, 2\pi]$. Contrary to the amplitude study presented here, a phase distribution test would not be affected by instrumental noise suppressing the target signal.
We foresee this as an additional valuable probe of the foreground Gaussianity at frequencies inaccessible by our test.
A detailed joint study of amplitude and phase foreground properties is crucial for the correct interpretation of the LISA data and is essential for an unbiased estimation of individual source parameters and the subsequent population inference.

\begin{acknowledgement}

The authors are grateful to E.~Finch and F.~Pozzoli for valuable inputs and N.~Karnesis for stimulating conversations and providing datasets in support of this study.
R.B. acknowledges support through the Italian Space Agency grant \emph{Phase A activity for LISA mission, n. 2017--29--H.0}.
A.K. acknowledges support of the UK Space Agency grant, no. ST/V002813/1.
R.B. and D.G. are supported by MUR Grant ``Progetto Dipartimenti di Eccellenza 2023-2027'' (BiCoQ), and by the ICSC National Research Center funded by NextGenerationEU.
D.G. is supported by
ERC Starting Grant No.~945155--GWmining, Cariplo Foundation Grant No.~2021-0555,
MUR PRIN Grant No.~2022-Z9X4XS, and MSCA Fellowships No.~101064542--StochRewind and No.~101149270--ProtoBH.
Computational work was performed using University of Birmingham BlueBEAR High Performance Computing facility and CINECA with allocations through INFN, Bicocca, and ISCRA project HP10BEQ9JB.
\textit{Software}:
We acknowledge usage of 
\textsc{Mathematica}~\cite{Mathematica} 
and of the following 
\textsc{Python}~\cite{python} 
packages for modeling, analysis, post-processing, and production of results throughout:
\textsc{matplotlib}~\cite{2007CSE.....9...90H},
\textsc{numpy}~\cite{2020Natur.585..357H},
\textsc{scipy}~\cite{2020NatMe..17..261V}.

\noindent\textbf{Data Availability Statement} The datasets generated during and/or analysed during the current study are available from the corresponding author on reasonable request.

\noindent\textbf{Code Availability Statement} The code/software generated during and/or analysed
during the current study is available from the corresponding author on reasonable request.

\noindent\textbf{Open Access} This article is licensed under a Creative Commons Attribution 4.0 International License, which permits use, sharing, adaptation, distribution and reproduction in any medium or format, as long as you give appropriate credit to the original author(s) and the source, provide a link to the Creative Commons licence, and indicate if changes were made. The images or other third party material in this article are included in the article’s Creative Commons licence, unless indicated otherwise in a credit line to the material. If material is not included in the article’s Creative Commons licence and your intended use is not permitted by statutory regulation or exceeds the permitted use, you will need to obtain permission directly from the copyright holder. To view a copy of this licence, visit \href{http://creativecommons.org/licenses/by/4.0/}{http://creativecommons.org/licenses/by/4.0/}.
Funded by $\textrm{SCOAP}^3$.
\end{acknowledgement}

\clearpage
\bibliographystyle{spphys}       %
\bibliography{main}   %

\appendix
\section{Stochastic signals envelopes}
\label{app:envelopes}
In order to derive the envelope of the LISA response to unresolved Galactic sources, we start by writing the LISA response to a single binary source in the low-frequency approximation. The spacecraft orbits can be described by the following Keplerian orbits:
\begin{align}
\bm{P}_j &= r_j \left(
\begin{array}{c}
\cos i \cos \beta_j \cos v_j - \sin\beta_j \sin v_j\\
\cos i \sin \beta_j \cos v_j + \cos \beta_j \sin v_j\\
- \sin i \cos v_j
\end{array}
\right), \label{eq:positions} \\
r_j &= \frac{R \left( 1 - e^2 \right)}{1 + e \cos v_j}, \\
\beta_j &= \frac{2 \pi j}{3} + \beta_0,
\end{align}
where $\bm{P}_j$ is the position of spacraft $j \in \{1,2,3\}$ with semimajor axis $R = 1~\text{AU}$, true anomaly $v_j$, inclination $i$, and eccentricity $e$.

The mean anomaly $l_j$ increases linearly with time $t$, and is related to the true anomaly by
\begin{align}
l_j &= \Omega t + \alpha_0 - \beta_j = u_j - e \sin u_j, \label{eq:Keplereq} \\
\tan \frac{u_j}{2} &= \sqrt{\frac{1 - e}{1+e}} \tan \frac{v_j}{2},
\end{align}
with mean orbital angular frequency $\Omega = 2\pi/\text{yr}$, mean anomaly $l_j$, and eccentric anomaly $u_j$.

In the following, we will use $\alpha_0 = \beta_0 = 0$ to simplify the derivation. The signal in the low frequency approximation from a binary with sky angles $(b_S, l_S)$, with an arbitrary $\alpha_0$ and $\beta_0$, can then be computed with
\begin{align}
 h_A(b_S, l_S) &= \cos 2 (\beta_0 - \alpha_0) h_{A,0}(b_S, l_S - \alpha_0) \nonumber\\
& - \sin 2 (\beta_0 - \alpha_0)  h_{E,0}(b_S, l_S - \alpha_0), \\
 h_E(b_S, l_S) &= \sin 2 (\beta_0 - \alpha_0) h_{A,0}(b_S, l_S - \alpha_0) \nonumber\\
&+ \cos 2 (\beta_0 - \alpha_0)  h_{E,0}(b_S, l_S - \alpha_0),
\end{align}
where $h_{A,0}$ and $h_{E,0}$ have been computed assuming $\alpha_0 = \beta_0 = 0$.

\subsection{Response to an individual DWD\label{app:individualenvelope}}

We assume that the GW signal is described in the Solar System barycenter by
\begin{align}
h_{ab}(t) &= - 2 A \left( 1 + \cos^2 \iota \right) e_{ab}^+ \cos 2 \phi(t) \nonumber\\
&+ 4 A \cos \iota \ e_{ab}^\times \sin 2 \phi(t), \\
\phi &= \omega_0 t + \phi_0, \\
e_{ab}^+ &= \epsilon_{ab}^+ \cos 2 \psi - \epsilon_{ab}^\times \sin 2 \psi
, \\
e_{ab}^\times &= \epsilon_{ab}^+ \sin 2 \psi + \epsilon_{ab}^\times \cos 2 \psi, \\
\epsilon_{ab}^+ &= \hat{p}^a \hat{p}^b - \hat{q}^a \hat{q}^b, \\
\epsilon_{ab}^\times &= \hat{p}^a \hat{q}^b + \hat{p}^b \hat{q}^a, \\
\uvec{p} &= \left( \sin b_S \cos l_S, \sin b_S \sin l_S, - \cos b_S \right), \\
\uvec{q} &= \left( \sin l_S, - \cos l_S, 0 \right).
\end{align}
The GW signal is described by the following parameters
\begin{itemize}
\item $A$, the amplitude
\item $\iota$, the inclination
\item $\omega_0$, the orbital angular frequency
\item $\phi_0$, the initial orbital phase
\item $\psi$, a polarization angle
\item $(b_S, l_S)$, the ecliptic coordinates of the source's sky location
\end{itemize}

In the low-frequency approximation, the LISA response to this wave can be modelled by two noise-independent GW detectors with signals $h_A$ and $h_E$ described by
\begin{align}
h_A(t) &= \frac{1}{\sqrt{2}} \left[ h_Z(t) - h_X(t) \right], \\
h_E(t) &= \frac{1}{\sqrt{6}} \left[ h_X(t) - 2 h_Y(t) + h_Z(t) \right], \\
h_X(t) &= \left[ \hat{L}_3^a(t) \hat{L}_3^b(t) - \hat{L}_2^a(t) \hat{L}_2^b(t) \right] h_{ab}(u), \\
h_Y(t) &= \left[ \hat{L}_1^a(t) \hat{L}_1^b(t) - \hat{L}_3^a(t) \hat{L}_3^b(t) \right] h_{ab}(u), \\
h_Z(t) &= \left[ \hat{L}_2^a(t) \hat{L}_2^b(t) - \hat{L}_1^a(t) \hat{L}_1^b(t) \right] h_{ab}(u), \\
u &= t - \uvec{k} \cdot \uvec{P}(t), \\
\uvec{k} &= \left( - \cos b_S \cos l_S , - \cos b_S \sin l_S, - \sin b_S \right), \\
\uvec{P} (t) &= R \left( \cos \Omega t , \sin \Omega t, 0 \right),
\end{align}
where $R = 1 \text{~AU}$ is the distance from the Sun to the detector barycenter, $\Omega = 2\pi/\text{yr}$ is the detector barycenter angular orbital frequency, and the directions of the LISA arms $\uvec{L}_i(t)$ can be described by
\begin{align}
\uvec{L}_1 &= - \frac{1}{2} \uvec{x} - \frac{\sqrt{3}}{2} \uvec{y}, \\
\uvec{L}_2 &= \uvec{x}, \\
\uvec{L}_3 &= - \frac{1}{2} \uvec{x} + \frac{\sqrt{3}}{2} \uvec{y},
\end{align}
where $\uvec{x}$ and $\uvec{y}$ are part of a triad tied to the detector arms together with $\uvec{z}$, which are expressed in a fixed ecliptic frame by
\begin{align}
\uvec{x} &= \left[ \frac{1}{4} (3 - \cos 2 \Omega t), - \frac{1}{4} \sin 2 \Omega t, \frac{\sqrt{3}}{2} \cos \Omega t \right], \\
\uvec{y} &= \left[ -\frac{1}{4} \sin 2 \Omega t, \frac{1}{4} (3 + \cos 2 \Omega t), \frac{\sqrt{3}}{2} \sin \Omega t \right], \\
\uvec{z} &= \left[ - \frac{\sqrt{3}}{2} \cos \Omega t, - \frac{\sqrt{3}}{2} \sin \Omega t, \frac{1}{2} \right].
\end{align}

Note that the signals in the usual noise-independent low-frequency LISA detectors are $h_I = h_A / \sqrt{6}$ and $h_{II} = - h_E / \sqrt{6}$.

In order to model the envelope of a signal constituting of the sum of a large number of signals, we start by computing the inclination and polarization averages of the signal from an individual source.

The structure of $h_A$ and $h_E$ can be expressed by setting $A=1$,
\begin{align}
 h_{A,E}(t) &= 2 C_{A,E}(t) \left( 1 + \cos^2 \iota \right) \cos 2 \phi(t) \nonumber \\
 &+ 4 S_{A,E}(t) \cos \iota \ \sin 2 \phi(t).
\end{align}
This allows us to compute the inclination and orbital averages
\begin{align}
 \left\langle h_{A,E} \right\rangle_{(\iota)}^2 &= \frac{1}{2} \int_{-1}^1 d\cos\iota \ \frac{1}{T} \int_0^T dt h_{A,E}(t)^2 \\
&= \frac{56}{15} C_{A,E}(t)^2 + \frac{8}{3} S_{A,E}(t)^2.
\end{align}
We can then compute the polarization averages
\begin{align}
\left\langle h_{A,E} \right\rangle_{(\iota,\psi)}^2 &= \frac{1}{\pi} \int_0^\pi d\psi  \left\langle h_{A,E} \right\rangle_{(\iota)}^2 \\
&= \sum_{n=-8}^8 \Big[ H_{(c;A,E;\iota,\psi;n)}^2 \cos n \Omega t \nonumber\\
&+ H_{(s;A,E;\iota,\psi;n)}^2 \sin n \Omega t \Big].
\end{align}
We can write the result using 
\begin{align}
 h_S^2 &= \left\langle h_{A} \right\rangle_{(\iota,\psi)}^2 + \left\langle h_{E} \right\rangle_{(\iota,\psi)}^2 \\
&= \sum_{n = 0}^4 h_{S;(n)}^2 (b_S) \cos n \Delta l_L, \\
 h_D^2 &= \left\langle h_{A} \right\rangle_{(\iota,\psi)}^2 - \left\langle h_{E} \right\rangle_{(\iota,\psi)}^2 \\
&= \sum_{n = 0}^8 h_{D;(n)}^2 (b_S) \cos \left( n \Delta l_L + 4 \bar{l}_S \right), \\
 \Delta l_L &= \Omega t - l_S , \\
\bar{l}_S &= l_S + \frac{\pi}{12},
\end{align}
with coefficients given by
\begin{align}
h_{S;(0)}^2 &= \frac{9}{320} \left( 328 + 152 \cos^2 b_S - 37 \cos^4  b_S \right), \\
h_{S;(1)}^2 &= -\frac{9\sqrt{3}}{40} \cos b_S 
\sin b_S \left( 52 + 5 \cos^2 b_S \right), \\
h_{S;(2)}^2 &= \frac{81}{80} \cos^2 b_S \left( 10 - \cos^2 b_S \right), \\
h_{S;(3)}^2 &= - \frac{27\sqrt{3}}{40} \sin b_S \cos^3 b_S , \\
h_{S;(4)}^2 &= \frac{81}{320} \cos^4  b_S, \\
h_{D;(4)}^2 &= -\frac{81}{320} \left( 8 - 40 \cos^2  b_S + 35 \cos^4  b_S \right), \\
h_{D;(3)}^2 &= -\frac{81\sqrt{3}}{80} \sin  b_S \cos  b_S \left( 4 - 7 \cos^2 b_S \right), \\
h_{D;(5)}^2 &= \frac{27\sqrt{3}}{80} \sin b_S \cos b_S \left( 4 - 7 \cos^2 b_S \right), \\
h_{D;(2)}^2 &= -\frac{243}{160} \cos^2  b_S \left( 6 - 7 \cos^2  b_S \right), \\
h_{D;(6)}^2 &= -\frac{27}{160} \cos^2 b_S \left( 6 - 7 \cos^2  b_S \right), \\
h_{D;(1)}^2 &= -\frac{243\sqrt{3}}{80} \sin b_S \cos^3 b_S, \\
h_{D;(7)}^2 &= \frac{9\sqrt{3}}{80} \sin b_S \cos^3 b_S, \\
h_{D;(0)}^2 &= -\frac{729}{640} \cos^4 b_S, \\
h_{D;(8)}^2 &= -\frac{9}{640} \cos^4 b_S.
\end{align}

\subsection{Sky Distribution \label{app:skydist}}

Having computed the response averaged over the orbit, polarization and inclination angles, we can now compute the average response to a large number of sources located according to a certain distribution over the sky. Keeping in mind that we wish to describe the contribution from unresolved galactic binaries across the sky, we can make the ansatz that the sources are distributed according to the following distribution in ecliptic coordinates:

\begin{align}
\!p(b_S, l_S) =& \frac{1}{2\pi \sigma_1 \sigma_2} \exp \left[ - \frac{1}{2} \bm{\theta}^T \mathbb{M} \bm{\theta} \right], \\
\!\bm{\theta} =& \left( \begin{array}{c}
b_S - b_M \\
l_S - l_M \end{array} \right), \\
\mathbb{M} =&\! \left( \begin{array}{c c}
\cos \delta & \sin \delta \\
-\sin \delta & \cos \delta \end{array} \right)\!\!
\left( \begin{array}{c c}
\sigma_1^{-2} & 0 \\
0 & \sigma_2^{-2} \end{array} \right)\!\!
\left( \begin{array}{c c}
\cos \delta & -\sin \delta \\
\sin \delta & \cos \delta \end{array} \right),
\end{align}
corresponding to a bivariate Gaussian distribution centered at $(b_M, l_M)$ (the location of the Galactic center) rotated by an angle $\delta$ relative to the ecliptic plane. Note that the distribution is unrealistic in several ways: galactic binaries are unlikely to follow a bivariate Gaussian distribution in the sky, and the probability distribution is normalized over $(b_S, l_S) \in \mathbb{R}^2$ instead of the sky $(b_S, l_S) \in [-\pi/2, \pi/2] \times [0, 2\pi]$. 
However with a reasonable choice of parameters the severity of these problems can be mitigated, and we hope to reach a reasonable approximation to the Galactic foreground in the end.
What is left to accomplish at this stage is to compute the foreground model for the response to unresolved DWDs:
\vspace{1cm}

\begin{widetext}
\begin{equation}
H_{S,D}^2 (t, b_S, l_S, \sigma_1, \sigma_2, \delta) = \int_\mathbb{R} \cos b_S d b_S \int_\mathbb{R} d l_S \ p(b_S, l_S) h_{S,D}^2,
\end{equation}
where for simplicity we integrate over the whole domain of the probability distribution rather than just the sky.
In order to carry out this integral, we can make use of the following result:
\begin{align}
\int_\mathbb{R} d b_S \int_\mathbb{R} d l_S \ p( b_S,  l_S) e^{i n  b_S} e^{i m  l_S}\\
= \exp \left[ - \frac{1}{4} \left( m^2 + n^2 \right) \left( \sigma_1^2 + \sigma_2^2 \right) + \frac{1}{4}  \left( m^2 - n^2 \right) \left( \sigma_1^2 - \sigma_2^2 \right) \cos 2 \delta + \frac{m n}{2} \left( \sigma_1^2 - \sigma_2^2 \right) \sin 2 \delta \right] e^{i n b_M} e^{i m l_M}.
\end{align}

To facilitate the presentation of the results we decompose $H^2_{S,D}$ as follows
\begin{align}
H_{S}^2 &= \frac{1}{10240} \sum_{n=0}^4 H_{S;(n)}^2, \\
H_{D}^2 &= \frac{1}{10240} \sum_{n=0}^8 H_{D;(n)}^2 ,
\end{align}
and define
\begin{align}
 \sigma_S^2 &= \sigma_1^2 + \sigma_2^2, \\
 \sigma_D^2 &= \sigma_1^2 - \sigma_2^2, \\
 \sigma_c^2 &= \sigma_D^2 \cos 2 \delta, \\
 \sigma_s^2 &= \sigma_D^2 \sin 2 \delta, \\
 \sigma_+^2 &= \sigma_S^2 + \sigma_c^2, \\
\bar{l}_M &= l_M + \frac{\pi}{12}, \\
\Delta \phi_L &= \Omega t - l_M.
\end{align}
Overall, we obtain for each term:
\begin{align}
H_{S;(0)}^2 &= e^{-\frac{1}{4} \sigma_S^2 - \frac{1}{4} \sigma_c^2 } \left( 120636 \cos b_M + 7614 e^{-2 \sigma_+^2} \cos 3 b_M - 666 e^{-6 \sigma_+^2} \cos 5 b_M \right) , 
 \end{align}

\begin{align}
H_{S;(1)}^2 &= \sqrt{3} \ e^{-\frac{1}{2}\sigma_S^2} \bigg[ -31392 \left( \cosh \frac{\sigma_s^2}{2} \sin b_M \cos \Delta \phi_L + \sinh \frac{\sigma_s^2}{2} \cos b_M \sin \Delta \phi_L \right) \nonumber\\
&- 32112 e^{-2 \sigma_+^2} \left( \cosh \frac{3\sigma_s^2}{2} \sin 3 b_M \cos \Delta \phi_L + \sinh \frac{3\sigma_s^2}{2} \cos 3 b_M \sin \Delta \phi_L \right) \nonumber\\
&- 720 e^{-6 \sigma_+^2} \left( \cosh \frac{5\sigma_s^2}{2} \sin 5 b_M \cos \Delta \phi_L + \sinh \frac{5\sigma_s^2}{2} \cos 5 b_M \sin \Delta \phi_L \right)  \bigg],
\end{align}

\begin{align}
H_{S;(2)}^2 &= e^{-\frac{5}{4} \sigma_S^2 + \frac{3}{4} \sigma_{c}^2} \Big[ 71280 \left( \cosh \sigma_s^2 \cos b_M \cos 2 \Delta \phi_L - \sinh \sigma_s^2 \sin b_M \sin 2 \Delta \phi_L \right) \nonumber\\
&+ 22680 e^{-2 \sigma_+^2} \left( \cosh 3 \sigma_s^2 \cos 3 b_M \cos 2 \Delta \phi_L - \sinh 3 \sigma_s^2 \sin 3 b_M \sin 2 \Delta \phi_L \right) \nonumber\\
&- 648 e^{-6 \sigma_+^2} \left( \cosh 5 \sigma_s^2 \cos 5 b_M \cos 2 \Delta \phi_L - \sinh 5 \sigma_s^2 \sin 5 b_M \sin 2 \Delta \phi_L \right)  \Big],
\end{align}

\begin{align}
H_{S;(3)}^2 &= \sqrt{3} \ e^{-\frac{5}{2}\sigma_S^2 + 2 \sigma_{c}^2} \bigg[ -864 \left( \cosh \frac{3\sigma_s^2}{2} \sin b_M \cos 3 \Delta \phi_L + \sinh \frac{3 \sigma_s^2}{2} \cos b_M \sin 3 \Delta \phi_L \right) \nonumber\\
&- 1296 e^{-2 \sigma_+^2} \left( \cosh \frac{9 \sigma_s^2}{2} \sin 3 b_M \cos 3 \Delta \phi_L + \sinh \frac{9\sigma_s^2}{2} \cos 3 b_M \sin \Delta 3 \phi_L \right) \nonumber\\
&- 432 e^{-6 \sigma_+^2} \left( \cosh \frac{15\sigma_s^2}{2} \sin 5 b_M \cos 3 \Delta \phi_L + \sinh \frac{15\sigma_s^2}{2} \cos 5 b_M \sin \Delta 3 \phi_L \right)  \bigg],
\end{align}

\begin{align}
H_{S;(4)}^2 &= e^{-\frac{17}{4} \sigma_S^2 + \frac{15}{4} \sigma_{c}^2} \Big[ 1620 \left( \cosh 2 \sigma_s^2 \cos b_M \cos 4 \Delta \phi_L - \sinh 2 \sigma_s^2 \sin b_M \sin 4 \Delta \phi_L \right) \nonumber\\
&+ 810 e^{-2 \sigma_+^2} \left( \cosh 6 \sigma_s^2 \cos 3 b_M \cos 4 \Delta \phi_L - \sinh 6 \sigma_s^2 \sin 3 b_M \sin 4 \Delta \phi_L \right) \nonumber\\
&+162 e^{-6 \sigma_+^2} \left( \cosh 10 \sigma_s^2 \cos 5 b_M \cos 4 \Delta \phi_L - \sinh 10 \sigma_s^2 \sin 5 b_M \sin 4 \Delta \phi_L \right)  \Big],
\end{align}
and
\begin{align}
H_{D;(4)}^2 &= e^{-\frac{1}{4} \sigma_S^2 - \frac{1}{4} \sigma_{c}^2} \left( 324 \cos b_M - 240 e^{-2 \sigma_+^2} \cos 3 b_M
 - 5670 e^{-6 \sigma_+^2} \cos 5 b_M \right) \cos \left( 4 \phi_L + 4 \bar{l}_M \right),
 \end{align}

\begin{align}
H_{D;(3)}^2 &= \sqrt{3} \ e^{-\frac{1}{2} \sigma_S^2} \bigg\{ -1296 \left[ \cosh \frac{\sigma_s^2}{2} \sin b_M \cos (3 \Delta \phi_L + 4 \bar{l}_M) - \sinh \frac{\sigma_s^2}{2} \cos b_M \sin (3 \Delta \phi_L + 4 \bar{l}_M) \right] \nonumber\\
&+3240 e^{-2 \sigma_+^2}  \left[ \cosh \frac{3\sigma_s^2}{2} \sin 3 b_M \cos (3 \Delta \phi_L + 4 \bar{l}_M) - \sinh \frac{3\sigma_s^2}{2} \cos 3 b_M \sin (3 \Delta \phi_L + 4 \bar{l}_M) \right] \nonumber\\
&+4536 e^{-6 \sigma_+^2} \left[ \cosh \frac{5\sigma_s^2}{2} \sin 5 b_M \cos (3 \Delta \phi_L + 4 \bar{l}_M) - \sinh \frac{5 \sigma_s^2}{2} \cos 5 b_M \sin (3 \Delta \phi_L + 4 \bar{l}_M) \right]  \bigg\},
\end{align}

\begin{align}
H_{D;(5)}^2 &= \sqrt{3} \ e^{-\frac{1}{2} \sigma_S^2} \bigg\{ 432 \left[ \cosh \frac{\sigma_s^2}{2} \sin b_M \cos (5 \Delta \phi_L + 4 \bar{l}_M) + \sinh \frac{\sigma_s^2}{2} \cos b_M \sin (5 \Delta \phi_L + 4 \bar{l}_M) \right] \nonumber\\
&-1080 e^{-2 \sigma_+^2} \left[ \cosh \frac{3\sigma_s^2}{2} \sin 3 b_M \cos (5 \Delta \phi_L + 4 \bar{l}_M) + \sinh \frac{3\sigma_s^2}{2} \cos 3 b_M \sin (5 \Delta \phi_L + 4 \bar{l}_M) \right] \nonumber\\
&-1512 e^{-6 \sigma_+^2} \left[ \cosh \frac{5\sigma_s^2}{2} \sin 5 b_M \cos (5 \Delta \phi_L + 4 \bar{l}_M) + \sinh \frac{5 \sigma_s^2}{2} \cos 5 b_M \sin (5 \Delta \phi_L + 4 \bar{l}_M) \right]  \bigg\},
\end{align}

\begin{align}
H_{D;(2)}^2 &= e^{-\frac{5}{4} \sigma_S^2 + \frac{3}{4} \sigma_{c}^2} \Big\{ -1944 \left[ \cosh \sigma_s^2 \cos b_M \cos (2 \Delta \phi_L + 4 \bar{l}_M) + \sinh \sigma_s^2 \sin b_M \sin (2 \Delta \phi_L + 4 \bar{l}_M) \right] \nonumber\\
&+ 10692 e^{-2 \sigma_+^2} \left[ \cosh 3 \sigma_s^2 \cos 3 b_M \cos (2 \Delta \phi_L + 4 \bar{l}_M) + \sinh 3 \sigma_s^2 \sin 3 b_M \sin (2 \Delta \phi_L + 4 \bar{l}_M) \right] \nonumber\\
&+ 6804 e^{-6 \sigma_+^2} \left[ \cosh 5 \sigma_s^2 \cos 5 b_M \cos (2 \Delta \phi_L + 4 \bar{l}_M) + \sinh 5 \sigma_s^2 \sin 5 b_M \sin (2 \Delta \phi_L + 4 \bar{l}_M) \right]  \Big\}, 
\end{align}

\begin{align}
H_{D;(6)}^2 &=  e^{-\frac{5}{4} \sigma_S^2 + \frac{3}{4} \sigma_{c}^2} \Big\{ -216 \left[ \cosh \sigma_s^2 \cos b_M \cos (6 \Delta \phi_L + 4 \bar{l}_M) - \sinh \sigma_s^2 \sin b_M \sin (6 \Delta \phi_L + 4 \bar{l}_M) \right] \nonumber\\
&+ 1188 e^{-2 \sigma_+^2} \left[ \cosh 3 \sigma_s^2 \cos 3 b_M \cos (6 \Delta \phi_L + 4 \bar{l}_M) - \sinh 3 \sigma_s^2 \sin 3 b_M \sin (6 \Delta \phi_L + 4 \bar{l}_M) \right] \nonumber\\
&+ 756 e^{-6 \sigma_+^2} \left[ \cosh 5 \sigma_s^2 \cos 5 b_M \cos (6 \Delta \phi_L + 4 \bar{l}_M) - \sinh 5 \sigma_s^2 \sin 5 b_M \sin (6 \Delta \phi_L + 4 \bar{l}_M) \right]  \Big\},
\end{align}

\begin{align}
H_{D;(1)}^2 &=  \sqrt{3} \ e^{-\frac{5}{2} \sigma_S^2 + 2 \sigma_c^2} \bigg\{ -3888 \left[ \cosh \frac{3\sigma_s^2}{2} \sin b_M \cos (\Delta \phi_L + 4 \bar{l}_M) - \sinh \frac{3\sigma_s^2}{2} \cos b_M \sin (\Delta \phi_L + 4 \bar{l}_M) \right] \nonumber\\
&-5832 e^{-2 \sigma_+^2} \left[ \cosh \frac{9\sigma_s^2}{2} \sin 3 b_M \cos (\Delta \phi_L + 4 \bar{l}_M) - \sinh \frac{9\sigma_s^2}{2} \cos 3 b_M \sin (\Delta \phi_L + 4 \bar{l}_M) \right] \nonumber\\
&-1944 e^{-6 \sigma_+^2} \left[ \cosh \frac{15\sigma_s^2}{2} \sin 5 b_M \cos (\Delta \phi_L + 4 \bar{l}_M) - \sinh \frac{15 \sigma_s^2}{2} \cos 5 b_M \sin (\Delta \phi_L + 4 \bar{l}_M) \right]  \bigg\},
\end{align}

\begin{align}
H_{D;(7)}^2 &= \sqrt{3} \ e^{-\frac{5}{2} \sigma_S^2 + 2 \sigma_c^2} \bigg\{ 144 \left[ \cosh \frac{3\sigma_s^2}{2} \sin b_M \cos (7 \Delta \phi_L + 4 \bar{l}_M) + \sinh \frac{3\sigma_s^2}{2} \cos b_M \sin (7 \Delta \phi_L + 4 \bar{l}_M) \right] \nonumber\\
&+216 e^{-2 \sigma_+^2} \left[ \cosh \frac{9\sigma_s^2}{2} \sin 3 b_M \cos (7 \Delta \phi_L + 4 \bar{l}_M) + \sinh \frac{9\sigma_s^2}{2} \cos 3 b_M \sin (7 \Delta \phi_L + 4 \bar{l}_M) \right] \nonumber\\
&+72 e^{-6 \sigma_+^2}  \left[ \cosh \frac{15\sigma_s^2}{2} \sin 5 b_M \cos (7 \Delta \phi_L + 4 \bar{l}_M) + \sinh \frac{15 \sigma_s^2}{2} \cos 5 b_M \sin (7 \Delta \phi_L + 4 \bar{l}_M) \right]  \bigg\},
\end{align}

\begin{align}
H_{D;(0)}^2 &= e^{-\frac{17}{4} \sigma_S^2 + \frac{15}{4} \sigma_{c}^2} \Big[ -7290 \left( \cosh 2 \sigma_s^2 \cos b_M \cos 4 \bar{l}_M + \sinh 2 \sigma_s^2 \sin b_M \sin 4 \bar{l}_M \right) \nonumber\\
&- 3645 e^{-2 \sigma_+^2} \left( \cosh 6 \sigma_s^2 \cos 3 b_M \cos 4 \bar{l}_M + \sinh 6 \sigma_s^2 \sin 3 b_M \sin 4 \bar{l}_M \right) \nonumber\\
&- 729 e^{-6 \sigma_+^2} \left( \cosh 10 \sigma_s^2 \cos 5 b_M \cos 4 \bar{l}_M + \sinh 10 \sigma_s^2 \sin 5 b_M \sin 4 \bar{l}_M \right)  \Big], 
\end{align}

\begin{align}
H_{D;(8)}^2 &= e^{-\frac{17}{4} \sigma_S^2 + \frac{15}{4} \sigma_{c}^2} \Big\{ -90 \left[ \cosh 2 \sigma_s^2 \cos b_M \cos (8 \Delta \phi_L + 4 \bar{l}_M) - \sinh 2 \sigma_s^2 \sin b_M \sin (8 \Delta \phi_L + 4 \bar{l}_M) \right] \nonumber\\
&- 45 e^{-2 \sigma_+^2} \left[ \cosh 6 \sigma_s^2 \cos 3 b_M \cos (8 \Delta \phi_L + 4 \bar{l}_M) - \sinh 6 \sigma_s^2 \sin 3 b_M \sin (8 \Delta \phi_L + 4 \bar{l}_M) \right] \nonumber\\
&- 9 e^{-6 \sigma_+^2} \left[ \cosh 10 \sigma_s^2 \cos 5 b_M \cos (8 \Delta \phi_L + 4 \bar{l}_M) - \sinh 10 \sigma_s^2 \sin 5 b_M \sin (8 \Delta \phi_L + 4 \bar{l}_M) \right]  \Big\}, 
\end{align}
The envelope of the signal in the A and E channels can then be computed with
\begin{align}
 h_{A,0}^2 &= \frac{1}{2} \left( H_S^2 + H_D^2 \right), \\
 h_{E,0}^2 &= \frac{1}{2} \left( H_S^2 - H_D^2 \right).
\end{align}
\end{widetext}
\end{document}